%% file: main.tex
\documentclass[a4paper, oneside, twocolumn, notitlepage, 10pt]{extarticle_ecoc}
\usepackage{ecoc}
\usepackage{textcomp}
\usepackage{acronym}
\input{acronyms}

\makeatletter
\renewenvironment{thebibliography}[1]
{
  \section*{\refname}%
  \footnotesize
  \list{\@biblabel{\@arabic\c@enumiv}}%
  {\settowidth\labelwidth{\@biblabel{#1}}%
    \leftmargin\labelwidth
    \advance\leftmargin\labelsep
    \usecounter{enumiv}%
    \let\p@enumiv\@empty
  \renewcommand\theenumiv{\@arabic\c@enumiv}}%
\sloppy\clubpenalty4000\widowpenalty4000\sfcode`\.\@m}
{\endlist}
\makeatother

\begin{document}
\selectlanguage{english}

\title{Sea-Scan: High-Accuracy, ML-based Dark Vessel Detection and Localisation via Weakly Supervised DAS Monitoring} %Sea-Scan: Dark Vessel Detection and Localization via Weakly Supervised DAS Monitoring over 120 km Subsea Fiber}

\author{
  Tian Tian \textsuperscript{(1)},
  Agastya Raj \textsuperscript{(1)},
  Lara Flanagan \textsuperscript{(1)},
  John Kennedy \textsuperscript{(2)},
  Marco Ruffini\textsuperscript{(1)}
}

\maketitle
\noindent\makebox[0pt][l]{%
  \parbox{\textwidth}{\centering
    \vspace{-8mm}
    \makebox[\linewidth][c]{\makebox[1.2\textwidth][c]{\textsuperscript{(1)}School of Computer Science and Statistics, IRIS Research Group, ADAPT Research Centre, Trinity College Dublin, Ireland}}\\
    \makebox[\linewidth][c]{\makebox[1.2\textwidth][c]{\textsuperscript{(2)}School of Engineering, ADAPT Research Centre, Trinity College Dublin, Ireland}}\\
    \makebox[\linewidth][c]{\makebox[1.2\textwidth][c]{\href{mailto:rajag@tcd.ie}{\textcolor{blue}{tianti@tcd.ie}}}}}}\par

\renewcommand\footnotemark{}
\renewcommand\footnoterule{}

\begin{strip}
  \begin{ecoc_abstract}
    We present an ML-based vessel detection and localization system, trained with weak supervision from imperfect AIS labels, that achieves a 97.8\% detection rate at 1.98\% false-trigger rate, successfully identifies dark-vessel events from unlabeled data.
  \end{ecoc_abstract}
\end{strip}
\section{Introduction}
Dark vessels operating with disabled or absent \ac{AIS} transponders pose direct threats to critical subsea infrastructure, including fiber-optic cables carrying over 95\% of intercontinental data traffic~\cite{dark_activity}. Detecting these vessels requires sensing modalities independent of cooperative transponders. \ac{DAS} transforms existing subsea fiber into a continuous sensor array with meter-scale spatial resolution over hundreds of kilometers, capable of detecting vessel-radiated noise without additional hardware \cite{baird_ocean_2025, xenaki_overview_2025}. The key challenge is converting this raw sensing capability into reliable, automated detection at operational scale.

Prior \ac{DAS}-based vessel detection has followed two directions. Array-processing and propagation-modeling approaches \cite{rivet_preliminary_2021, landro_sensing_2022, paap_leveraging_2025} can localize vessels but require calibrated environmental parameters and expert tuning that limit scalability. Recent deep learning methods \cite{huang_daship_2025, ramirez-torres_vessel_2025, zhang_dasnet_2026,shao_tracking_2025} reduce manual configuration by training on AIS-aligned \ac{DAS} data, but face three limitations that restrict operational use. First, models are typically developed on short cable segments (a few to tens of kilometers)\cite{zhang_dasnet_2026,shao_tracking_2025}, where noise characteristics and seabed coupling conditions are relatively uniform; performance across the full heterogeneity of a long-range link remains undemonstrated. Second, detection is often simplified to binary presence/absence classification \cite{ramirez-torres_vessel_2025} or restricted to strong wake-crossing signatures \cite{huang_daship_2025}, precluding along-cable localization and early detection of approaching vessels. Third, and most critically, AIS-derived labels are treated as reliable ground truth despite systematic noise from timing and position misalignment, intermittent reporting, and AIS-silent targets, which biases both training and evaluation.

In this work, we present Sea-Scan, an ML-based end-to-end detection and localization framework, built on encoder backbone, designed to learn from imperfect AIS supervision at long-range scale. We report results from 35 days of continuous \ac{DAS} monitoring over a 120 km subsea cable in the Irish Sea. The main contributions of the proposed framework include: (i) a weakly supervised training objective combining top-K \ac{MIL} with temporal smoothness constraints which accommodates label noise; (ii) a factorized spatio-temporal detection head that separates temporal event detection from along-cable localization via multiplicative gating, suppressing false activations in the absence of segment-wide temporal evidence; and (iii) a hysteresis-based triggering mechanism with trend consistency filtering, converting continuous model outputs into stable detection alerts. On held-out test set spanning the full cable, the system achieves 97.8\% detection rate at 1.98\% false-trigger rate, with median along-cable localization error of 239.9 m relative to AIS-reported crossing positions. Applied to segments with no AIS-reported vessels, Sea-Scan identifies 42 candidate dark vessel transits with spectral signatures indicative of vessel radiated noise.

\begin{figure*}[!t]
    \centering
    \includegraphics[width=1.0\linewidth]{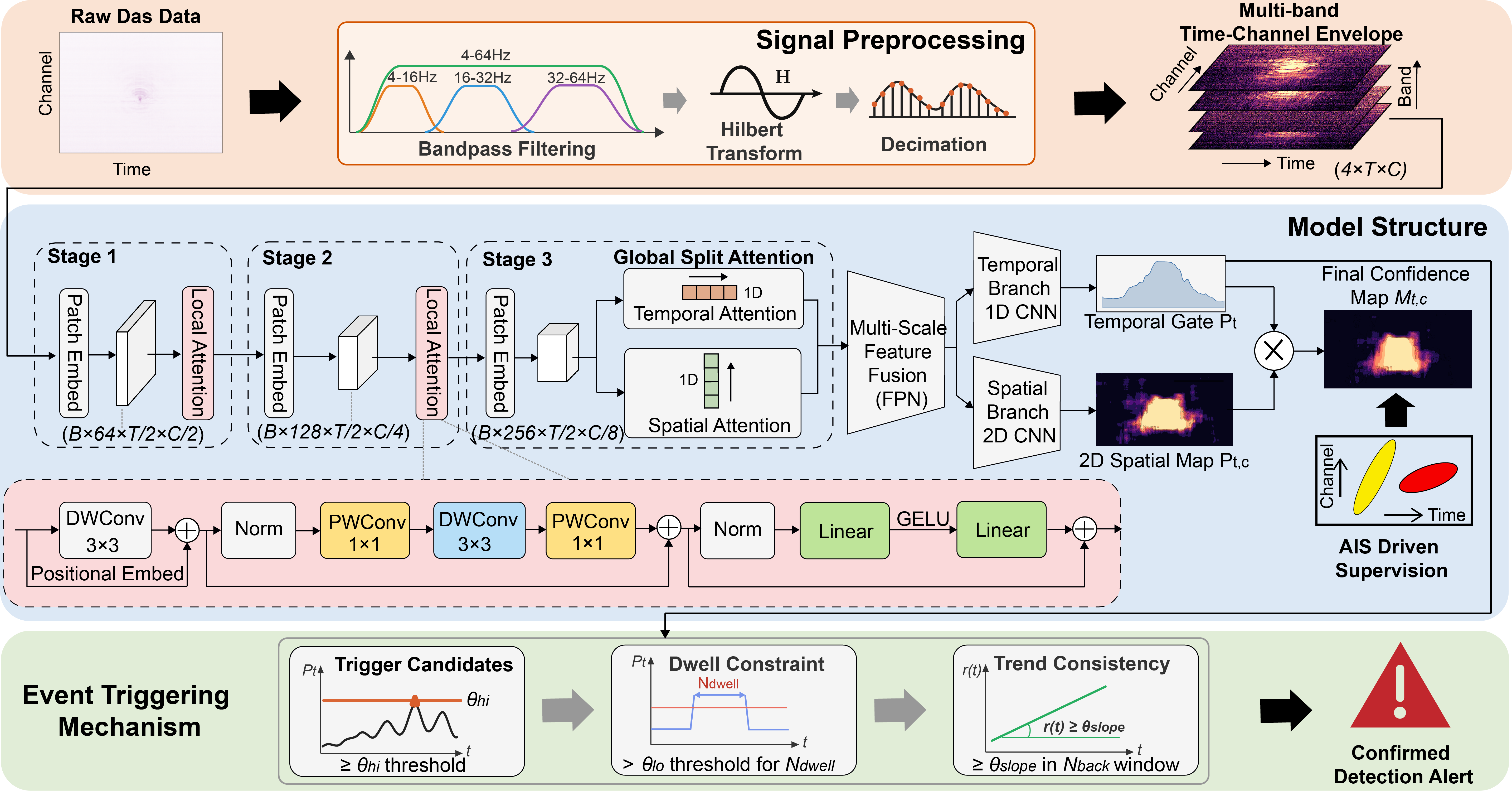}
    \vspace{-5mm}
    \caption{End-to-end pipeline of the proposed \ac{DAS} vessel detection and localization framework.}
    \label{fig:detection_framework}
\end{figure*}

\vspace{-3mm}
\section{Detection Framework}
\textbf{Preprocessing}. Fig.\ref{fig:detection_framework} summarizes the Sea-Scan pipeline from raw \ac{DAS} recordings to detection alerts.  The raw strain-rate signal is bandpass-filtered into three sub-bands (4–16, 16–32, 32–64 Hz) and one broadband channel (4–64 Hz). Vessel-radiated noise typically spans from below 10 Hz to beyond 1 kHz, but the 61.28 m gauge length in our deployment imposes spatial averaging that attenuates content above $\sim$60 Hz. Within this bandwidth, different vessels exhibit distinct spectral energy distributions across sub-bands, motivating the multi-band decomposition. For each band, the amplitude envelope is extracted via the Hilbert transform and decimated to 0.625 Hz, retaining only slowly-varying energy relevant to vessel transits on minute timescales. The resulting four-band tensor is tiled with 50\% overlap in time and channel, and each tile is z-score normalized per band to compensate for sensitivity and \ac{SNR} variation along the cable. For weak supervision, AIS position reports are interpolated and projected onto the cable to form a binary corridor mask: channels within $\pm$2 km are labeled positive. This radius accounts for uncertainty from projection error and lateral extent of seabed acoustic coupling. The mask is intentionally conservative, providing candidate-positive regions that the model must learn to refine.

\textbf{Model Structure.} The encoder backbone adopts the hierarchical three-stage spatio-temporal architecture \cite{li_uniformer_2022}, progressively downsampling time and channel dimensions through local attention in early stages and split temporal/spatial attention in the final stage. Multi-scale features are fused via a Feature Pyramid Network (FPN) to produce a dense feature map $\mathbf{z}_F\in\mathbb{R}^{B\times D_F\times T\times C}$. For detection, we decouple \textit{when} a vessel is present from \textit{where} along the cable it is located using two parallel branches. A temporal branch aggregates $\mathbf{z}_F$ over all channels via a 1D CNN to produce a per-timestep activity score $P_t \in [0,1]$, while a spatial branch applies a 2D CNN to produce a dense time–channel map $P_{t,c} \in [0,1]$. The final confidence map is formed by multiplicative gating: $M_{t,c} = P_t \cdot P_{t,c}$. This factorization reflects a physical prior: $P_t$ acts as a global gate, requiring sustained temporal evidence (e.g., the gradually increasing then decreasing energy characteristic of a vessel transit) before any spatial detection is expressed. Impulsive interference which might be spatially localized but lacks the sustained temporal structure is suppressed by the gating mechanism.

\textbf{Training Objective.} The loss combines three terms designed for noisy AIS supervision: (i) a top-K \ac{MIL} loss on $P_t$ within AIS-positive intervals, where only the top 10\% of timesteps per positive bag contribute gradients, preventing penalization of mislabeled portions of the corridor, (ii) a negative suppression with Huber temporal smoothing, and (iii) pixel-wise \ac{BCE} plus Dice on the time-channel map to calibrate along-cable localization.

\textbf{Event Triggering.} $P_t$ is converted to alerts via hysteresis thresholding: a candidate is opened when $P_t > \theta_{\rm{hi}}$, confirmed if $P_t$ remains above $\theta_{\rm{lo}}$ for $N_{\rm{dwell}}$ samples, and cleared when $P_t < \theta_{\rm{lo}}$. A trend consistency filter additionally requires that over a backward window of $N_{\rm{back}}$ samples, the fraction of positive increments exceeds $\theta_{\rm{slope}}$, rejecting impulsive interference that lacks the gradual energy increase characteristic of an approaching vessel.

\begin{figure*}[!t]
    \centering
    \includegraphics[width=1.0\linewidth]{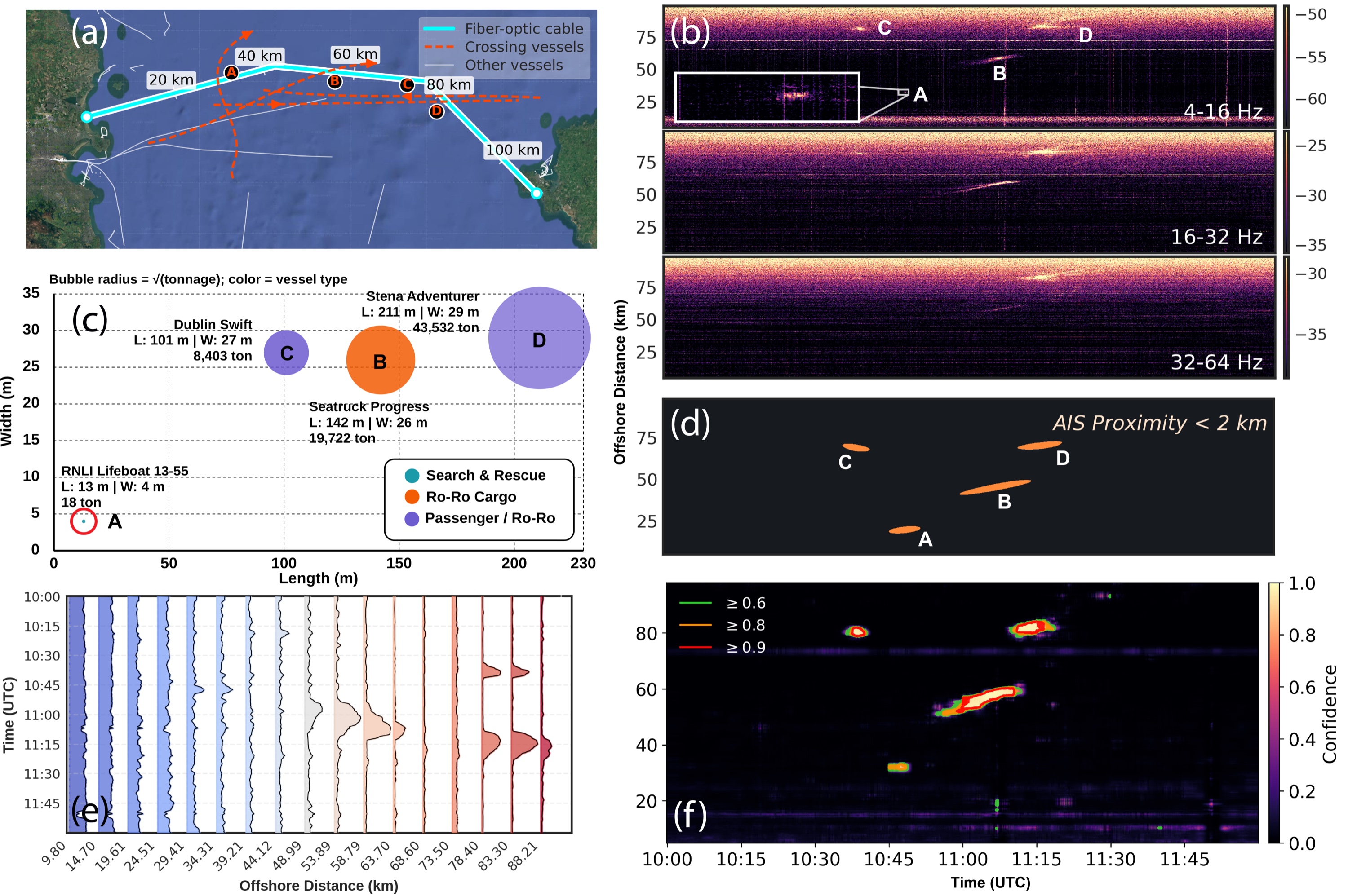}
    \vspace{-5mm}
    \caption{Detection example across 120 km of cable over a 2-hour interval. (a) EFBL cable route and AIS vessel trajectories. (b) Multi-band \ac{DAS} envelope intensity. (c) AIS-reported vessel characteristics for the four crossings. (d) AIS-derived mask of proximity corridor, projected on to the time-distance grid. (e) Temporal activity score $P_t$ for each 5 km cable segment. (f) Final confidence map $M_{t,c}$ with threshold contours.}
    \label{fig:detection-example}
\end{figure*}
\vspace{-3mm}
\section{Experimental Setup}  
\textcolor{black}{The experiment was conducted on the Emerald Fibre Bridge Link (EFBL), a UK–Ireland subsea route with a total fiber length of $\sim$120 km between Wales and Dublin, of which 12 km is on land. Unless stated otherwise, all distances in this work refer to offshore along-cable distance. \ac{DAS} measurements were acquired using an OptoDAS interrogator from Alcatel Submarine Networks (ASN), configured at 625 Hz sampling rate, 61.28 m gauge length, and 30.64 m channel spacing. Data were collected from 17 April to 22 May 2025 (35 days), totaling $\sim$14 TB of raw recordings. Using AIS-assisted screening, we identified 1,518 vessel-crossing events. Negative samples were extracted as 30-minute windows over 5 km cable sections with no AIS-reported vessels within $\pm 10$ km, yielding 4,343 noise events. The dataset was split into 70\% for training and 30\% for testing. During training, tiles were balanced to 70\% vessel and 30\% noise by randomly subsampling the noise class. Preprocessing and inference were performed on a single NVIDIA GeForce RTX 4090 GPU and a 13th Gen Intel\textsuperscript{\textregistered} Core\textsuperscript{\texttrademark} i9-13900K CPU, with 64 GB RAM.}
\vspace{-3mm}
\section{Results and Discussion}
 On the balanced test set of 454 vessel-crossing events and 454 noise events, the model achieves a detection rate of 97.8\% (TP=444, FN=10) and a false-trigger rate of 1.98\% (FP=9, TN=445). Of the 10 missed detections, 6 involved vessels at offshore distances exceeding 85 km, consistent with increased noise level from  at the far end of the cable. For localization, the predicted confidence map achieves a Dice score of 0.612 and IoU of 0.579 against the AIS-derived corridor mask. The peak-confidence location has a median offset of 239.9 m along the cable relative to AIS-reported crossing positions. 

Fig. \ref{fig:detection-example} reports, for demonstration purposes, an example of detection over the full offshore aperture by processing a two-hour interval (10:00–12:00 UTC, 2 May 2025). AIS recorded four crossings at offshore distances of 31.7, 57.4, 80.6, and 81.9 km, over that two-hour interval, spanning a 13 m lifeboat (38 ton) to a 211 m Ro-Ro carrier (43,532 ton). The temporal activity score $P_t$ per 5 km segment (Fig. \ref{fig:detection-example} (e)) and predicted confidence map $M_{t,c}$ (Fig. \ref{fig:detection-example} (f)) produce spatially coherent detections that coincide with elevated \ac{DAS} energy across all four crossings, including the small-vessel case with comparatively weak acoustic signatures. For a 30 min $\times$ 10 km tile, Sea-Scan requires 139 s of CPU preprocessing on average, and 0.238 s for GPU inference. This means a 10s time slot over 120km can be processed in 9.3 s, making the system capable of real time operation. 

To evaluate dark-vessel detection, we applied Sea-Scan to the full noise dataset (i.e., with no AIS information). The model flagged 42 of 4,343 noise events above the detection threshold. Manual review confirmed distinct tonal spectral components consistent with vessel-radiated noise. These candidate dark-vessel transits, undetectable by AIS-based monitoring, demonstrate the operational capability motivating this work.

\vspace{-2mm}
\section{Conclusion}
We showed how weakly supervised learning of DAS data that accommodates noisy AIS labels, can deliver automated, real time vessel detection and localization across a 120 km link, on commodity compute hardware. The system achieves 97.8\% detection at 1.98\% false-trigger rate with 239.9 m median localization error, and identified 42 candidate dark vessel transits in AIS-silent segments, confirmed by spectral signatures consistent with vessel radiated noise. %These results establish \ac{DAS}-based dark vessel detection as operationally viable for continuous subsea infrastructure monitoring.
\vspace{-2mm}
\section{Acknowledgments}
{{\small
This publication has emanated from research conducted with the financial support of Research Ireland and the Department of Defence under the Research Ireland-Department of Defence grant number 24/FIP/DO/13340P, in addition to 18/RI/5721 and 13/RC/2106 P2.
\par}}

%\clearpage

\end{document}

%% file: acronyms.tex
\acrodef{OLS}{Open Line System}
\acrodef{ROADM}{Reconfigurable Optical Add-Drop Multiplexer}
\acrodef{BVT}{Bandwidth Variable Transceiver}
\acrodef{OSNR}{Optical Signal to Noise Ratio}
\acrodef{ASE}{Amplified Spontaneous Emission}
\acrodef{NETCONF}{Network Configuration}
\acrodef{NLI}{Non Linear Interference}
\acrodef{BER}{Bit Error Rate}
\acrodef{OSaaS}{Optical Spectrum as a Service}
\acrodef{CFO}{Carrier Frequency Offset}
\acrodef{CDC}{Chromatic Dispersion Compensation}
\acrodef{DGD}{Differential Group Delay}
\acrodef{ORP}{Optical Received Power}
\acrodef{SNR}{Signal to Noise Ratio}
\acrodef{PDL}{Polarization Dependent Loss}
\acrodef{ML}{Machine Learning}
\acrodef{ANN}{Artificial Neural Network}
\acrodef{OCM}{Optical Channel Monitor}
\acrodef{OPM}{Optical Performance Monitoring}
\acrodef{EDFA}{Erbium Doped Fiber Amplifier}
\acrodef{EON}{Elastic Optical Network}
\acrodef{DWDM}{Dense Wavelength Division Multiplexing}
\acrodef{QoT}{Quality of Transmission}
\acrodef{WSS}{Wavelength Selective Switch}
\acrodef{OOK}{On-Off Keying}
\acrodef{PSD}{Power Spectral Density}
\acrodef{SLA}{Service Level Agreement}
\acrodef{XPM}{Cross Phase Modulation}
\acrodef{CPM}{Cross Polarization Modulation}
\acrodef{SOP}{State of Polarization}
\acrodef{PD}{Photodetector}
\acrodef{ILA}{In-Line Amplifier}
\acrodef{DAS}{Distributed Acoustic Sensing}
\acrodef{DL}{Deep Learning}
\acrodef{FPN}{Feature Pyramid Network}
\acrodef{MIL}{Multiple-Instance Learning}
\acrodef{BCE}{Binary Cross-Entropy}
\acrodef{ASN}{Alcatel Submarine Networks}
\acrodef{SNR}{Signal-to-Noise Ratio}
\acrodef{AIS}{Automatic Identification System}